\documentclass[]{iucr}    
\RequirePackage{float}
\journalcode{S}  

\usepackage{siunitx}
\usepackage{color}        
\usepackage[export]{adjustbox} 
\usepackage{multirow}
\usepackage{makecell}
\usepackage{wasysym}

\renewcommand{\deg}{$^{\circ}$}

\begin{document}

\title{Polished diamond x-ray lenses}

\author[a]{Rafael}{Celestre}
\author[b]{Sergey}{Antipov}\aufn{now at: Palm Scientific, 365 Remington blvd, Bolingbrook, IL, 60440 USA} 
\author[b]{Edgar}{Gomez}
\author[a]{Thomas}{Zinn}
\author[a]{Raymond}{Barrett}
\cauthor[a]{Thomas}{Roth}{troth@esrf.fr}{}

\aff[a]{ESRF - The European Synchrotron, 71 Avenue des Martyrs, 38000 Grenoble, \country{France}}
\aff[b]{Euclid Techlabs, 365 Remington blvd, Bolingbrook, IL, 60440 \country{USA}}


\maketitle  

\begin{synopsis}
Performance of state of the art diamond x-ray lenses.
\end{synopsis}


\begin{abstract}

We present high quality bi-concave 2D focusing diamond x-ray lenses of apex-radius $R=\SI{100}{\micro\meter}$ produced via laser-ablation and improved via a mechanical polishing process. Both for polished and unpolished individual lenses and for stacks of 10 lenses, we show the remaining figure errors determined using x-ray speckle tracking and compare these results to those of commercial $R=\SI{50}{\micro\meter}$ beryllium lenses that have similar focusing strength and physical aperture. For two stacks of 10 diamond lenses (polished and unpolished) and a stack of 11 beryllium lenses, we present measured 2D beam profiles out of focus and wire scans to obtain the beam size in the focal plane. These results are complemented with small angle x-ray scattering (SAXS) measurements of a polished and an unpolished diamond lens. Again, we compare this to the SAXS of a beryllium lens. The polished x-ray lenses show similar figure errors to commercially available Be lenses. While the beam size in the focal plane is comparable with that of the Be lenses, the SAXS signal of the polished diamond lenses is considerably lower.

\end{abstract}


\section{Introduction}\label{sec:intro}

Diamond is an excellent material for x-ray optics \cite{shvydko_diamond_2017}, as it can withstand high heat loads due to its unrivalled thermal conductivity, absorbs little due to its low atomic number ($Z$ = 6), can be obtained in pure and crystalline form with reasonable quality, and, as a refractive element such as a focusing lens, offers a  refraction-to-absorption ratio $\delta/\mu$ \cite{serebrennikov_optical_2016} higher than all typical lens materials except for Be.

Since the advent of x-ray focusing lenses \cite{tomie_x-ray_1997, snigirev_compound_1996}, significant effort has been put into the fabrication of diamond x-ray lenses. Due to the difficulty of machining diamond with conventional tools, initial trials deposited CVD diamond into moulds produced in silicon via semiconductor lithography techniques \cite{snigirev_diamond_2002} or used  direct etching of CVD diamond wafers \cite{nohammer_deep_2003,nohammer_diamond_2003}. These methods led to planar 1D focusing diamond lenses of reduced sagittal aperture ($\leq\SI{110}{\micro\meter}$) due to the difficulties of producing deep structures using these planar technologies. Despite later improvements in the processing  \cite{isakovic_diamond_2009,alianelli_planar_2010} lenses produced by these methods continue to display significant tilt of their side walls, i.e. they are not perpendicular to the wafer surface. Nanofocusing with planar diamond lenses was finally demonstrated by \cite{malik_deep_2013,fox_nanofocusing_2014,lyubomirskiy_diamond_2019}; however, the physical sagittal aperture of these nanofocusing lenses is still below $\sim$\SI{100}{\micro\meter}, and 2D focusing requires stacking of two 1D planar structures with consequent losses in transmission. Thicker 1D focusing structures were obtained by laser cutting of diamond plates \cite{polikarpov_diamond_2016,kononenko_fabrication_2016}, with a big challenge being again the verticality of the lens walls, reported to be 1.7\deg\ over a cutting depth of 0.6\,mm. 

More recently, with the emergence of pico- and femto-second pulsed lasers, fabrication of larger structures via laser ablation became viable \cite{terentyev_parabolic_2015,antipov_single-crystal_2016,terentyev_linear_2017}. Ablation allows the fabrication of 1D or 2D focusing lenses into bulk diamond. Since the as-fabricated surface is rough (rms roughness in the order of \SI{1}{\micro\meter} \cite{terentyev_parabolic_2015,jj-x-ray_single_2021}, the focusing properties of these lenses are worse than comparable commercial Be lenses \cite{lengeler_refractive_2012,lengeler_company_2021}. A post-processing step to improve the surface roughness is thus required. Different methods have been employed: mechanical polishing \cite{antipov_femtosecond_2018} and polishing via wet or dry etches \cite{jj-x-ray_single_2021}. Polishing using focused ion beam or excimer laser has also been proposed \cite{polikarpov_diamond_2016}. Focused ion beam produced diamond lenses have been shown to produce very smooth lens surfaces and lead to excellent focusing. However, in practice, slow fabrication speeds limit the application of such technologies to the production of few lenses of very small aperture \cite{medvedskaya_diamond_2020}. Similar sized lenses, used in the visible range and with larger radii can also be produced via a chemical reflow method \cite{zhu_fabrication_2017}.

We manufactured and characterised bi-concave 2D focusing diamond x-ray lenses produced, as individual elements, via laser ablation and subsequent mechanical polishing. The performance of these lenses is comparable to that of commercial Be lenses, which are regarded as the standard due to their wide-spread use and continuous development dating back to 2002 \cite{schroer_beryllium_2002}. The lenses were conditioned in \diameter~12\,mm frames, which conveniently makes them compatible with existing hardware (e.g. pin-holes, spacers, lens cases, v-blocks and transfocators).


\section{Lens fabrication}\label{sec:lens_fabrication}

\subsection{Ablation process}\label{sec:ablation}

Ultrafast (femtosecond) laser ablation of materials has attracted significant interest in recent years due to its promise for high machining accuracy and exceptional quality of the material processing. These appealing properties originate from laser pulse durations which are significantly shorter than the thermal diffusion time at the scale of the beam spot size, thus offering reduced thermal damage and efficient laser pulse utilisation \cite{cheng_review_2013,sugioka_ultrafast_2014}. Ultrafast laser ablation opened new opportunities for micro-machining materials like diamond which have a limited pool of processing technologies due to their physical properties. The diamond x-ray lenses characterised in this paper are produced by the femtosecond laser micro-machining \cite{osellame_femtosecond_2012}.  

Amongst the possible focusing shapes \cite{sanchez_del_rio_aspherical_2012},  2D-focusing x-ray lenses most commonly assume a bi-concave paraboloid geometry \cite{lengeler_imaging_1999}. When producing such a shape via laser ablation, we decompose this profile into circular layers with diameters that decrease parabolically with depth. In order to remove a circular layer, the laser beam is steered by motorised mirrors covering a circular area uniformly. 
To achieve the required accuracy of the shape a large set of parameters has to be optimised: the beam focus size and convergence as well as the laser flux and pulse duration. We utilise a green 515\,nm laser with a 200~fs pulse duration and average power of few hundred milliwatts depending on the focusing configuration. The unavoidable motion errors, triggering jitter and re-deposition of the ablated material result in figure errors - that is, deviations of the resultant profile from the design geometry. The lack of in-situ metrology makes it virtually impossible to iteratively eliminate correlated errors by adjusting the ablation recipe.

\subsection{Raw material}\label{sec:material}

The two main techniques for producing synthetic diamonds are high-pressure-high-temperature (HPHT) growth and the chemical vapour deposition (CVD) technique. While HPHT diamonds present better crystalline structure, CVD diamonds can be produced in larger sizes and are less expensive \cite{shvydko_diamond_2017}. Both types of synthetic diamond come in various grades, which differ in the growth parameters and in spurious nitrogen content. The lenses used in this study were structured in low dislocation density HPHT (100)-oriented diamond plates.

\begin{figure}
    \centering
    \includegraphics[width=\columnwidth]{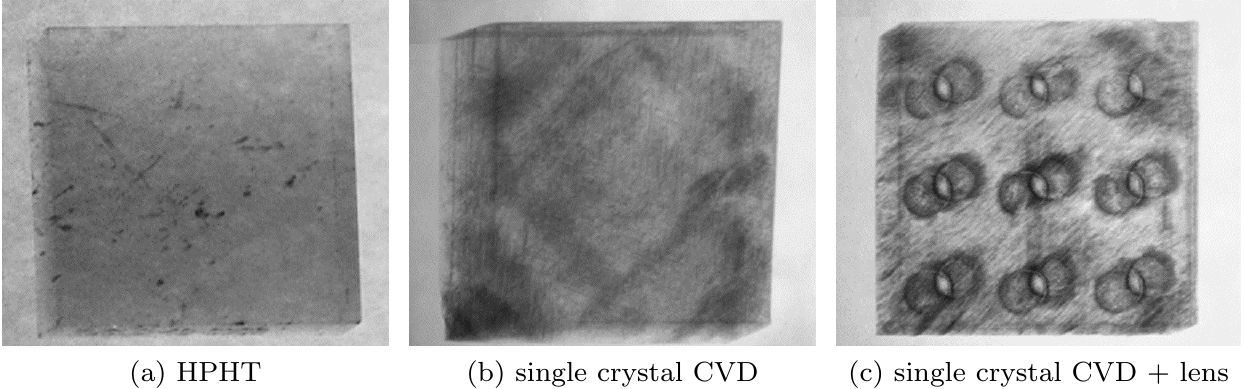}\vspace{-0.6cm}
    \caption{x-ray white beam topography images. (a) HPHT diamond, (b) single crystal CVD diamond and (c) CVD diamond plate with lenses cut in it. Diamonds in this image are of size $3\times3\times0.5$~mm$^3$.}
    \label{fig:topo}
\end{figure}

Figure~\ref{fig:topo} compares the white-beam topography for a HPHT and a CVD diamond. Due to lower dislocation densities in HPHT diamond as compared to CVD diamond, image (a) is more homogeneous than image (b). For the purposes of laser-cutting, we have not observed any substantial difference using several grades of diamonds, i.e. poly-crystalline and single-crystalline CVD and HPHT with different nitrogen concentration. The fs-laser cutting process does not significantly increase the dislocation density around the lens structures as shown in Fig.~\ref{fig:topo}(c). We have observed, however, a difference in the polishing process removal rate and uniformity depending on the diamond grade.  

In terms of x-ray transport and beam focusing, one expects that a diamond with higher degree of crystallinity will produce less small angle x-ray scattering (SAXS). However more important to the performance of a diamond lens are its figure error and surface roughness. The choice of diamond grade has a significant impact upon the total cost of the lens and if a lower grade diamond is acceptable for certain applications the cost of the diamond lenses can be reduced. Now that the quality of diamond lenses presented here is substantially improved, it is possible and desirable to perform a dedicated study on the influence of the raw material for future diamond lens production.

\subsection{Post-polishing}\label{sec:polishing}
Depending on ablation parameters, the as-cut surface roughness of the diamond lens is 300$-$500~nm S$_\textrm{a}$ (ISO 25178, which defines a surface extension of R$_\textrm{a}$, arithmetical mean deviation of heights from the assessed profile - ISO 4287), obtained by confocal microscopy at 50$\times$ magnification. To reduce these values we developed a post-ablation chemical–mechanical polishing procedure for the diamond lenses. In this process a conformal needle (polishing bit) is lowered into the diamond lens along with 0.1 micron diamond slurry and spun inside for several hours. 

\begin{figure}
    \centering
    \includegraphics[width=\columnwidth]{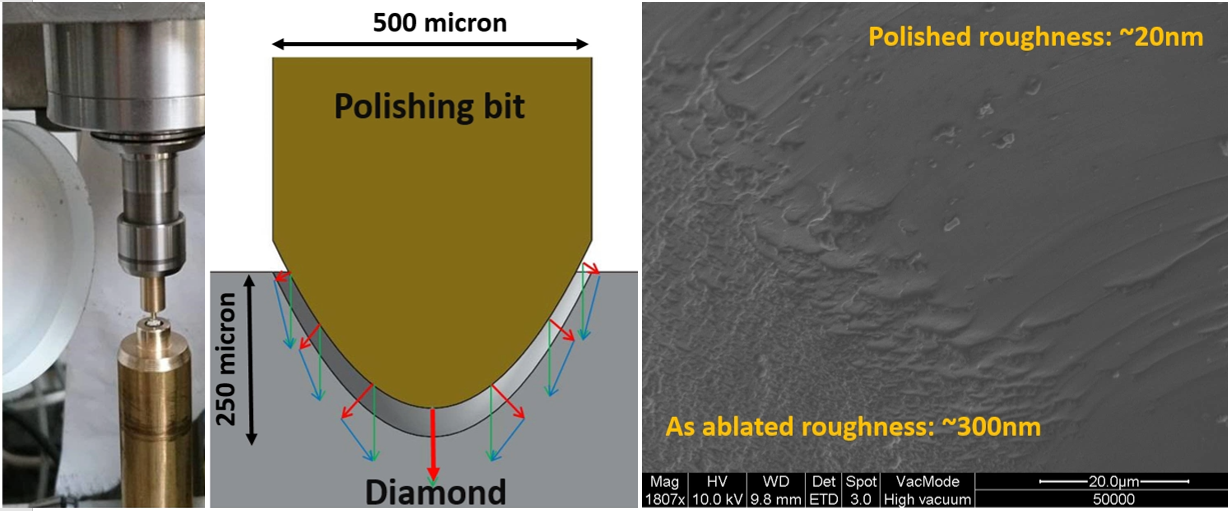}\vspace{-0.6cm}
    \caption{Picture and sketch of the polishing process (left and middle). On the right, an SEM image shows the diamond lens surface at an interface between a polished and intentionally unpolished region.}
    \label{fig:polishing}
\end{figure}

\begin{figure}
    \centering
    \includegraphics[width=0.7\columnwidth]{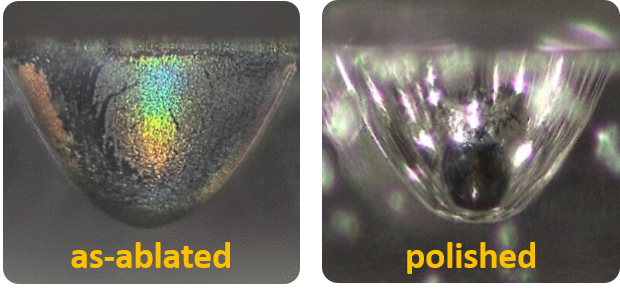}\vspace{-0.6cm}
    \caption{Microscope image of both an unpolished (left) and a polished (right) lens. The images are taken through the polished side wall of the diamond plate.}
    \label{fig:polishedUnpolished}
\end{figure}

Figure~\ref{fig:polishing} shows the basic setup and typical polishing results. A high speed, high accuracy spindle is used to spin the polishing bit inside the lens. Note that due to the small size of the lens and the polishing bit a microscope is used to align the polishing bit with the lens. The polishing process is a hybrid of chemical etching and free abrasive polishing. Multiple parameters must be optimised: polishing bit material; slurry solvent and grit size; contact pressure; and rotational speed. Since the surface of the lens is curved (Fig.~\ref{fig:polishing} - middle), there is an uneven force perpendicular to the surface when a downward pressure is applied. This leads to an uneven material removal. To compensate this effect we introduce periodic pressure applied sideways. The polishing process has been fine tuned and is able to maintain a quasi-uniform removal rate along the complete surface of the lens. The lens surface is polished to optical transparency with local micro-roughness of about 20\,nm S$_\textrm{a}$. The final result is presented in Fig.~\ref{fig:polishedUnpolished}, which was captured  through the polished and transparent side wall of the diamond plates. 

\subsection{Packaging}\label{sec:packaging}

In general, x-ray lenses are stacked together to obtain short focal lengths. These compound refractive lenses (CRL), as those stacks are called, can be composed of several dozens of individual lenses. An important factor to be considered for the adoption of CRLs in synchrotron beamlines is how well the lenses can be aligned with respect to each other as misalignments between the elements of a CRL will degrade the lens performance \cite{andrejczuk_role_2010, celestre_recent_2020}. A common solution applied to individually-produced 2D x-ray lenses is to house them in  \diameter~12\,mm precision-machined disks, where the lenses should be centred within 1 or \SI{2}{\micro\meter}. This is the case, e.g., for commercial Be, Al and Ni embossed lenses. 

\begin{figure}
    \centering
    \includegraphics[width=\columnwidth]{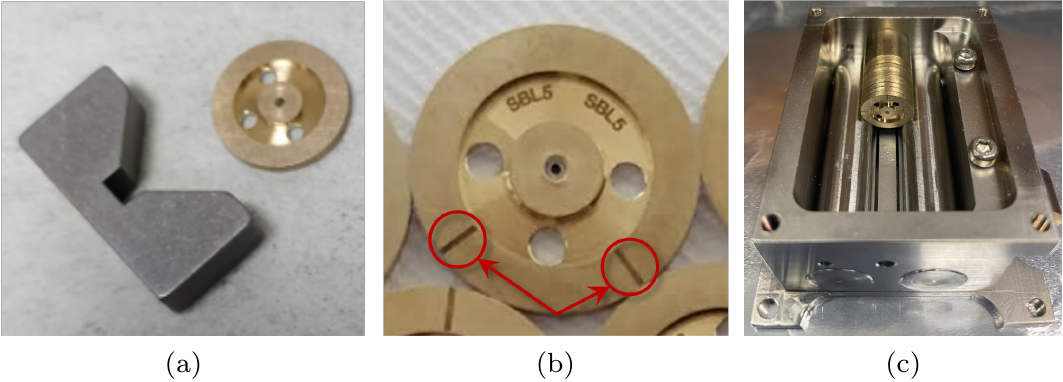}\vspace{-0.6cm}
    \caption{Packaging of diamond lens: a) mini v-block for lens ablation; b) lens support disk containing a diamond with ablated lens in the middle and fiducial markings indicating azimuthal orientation during the ablation process; c) lenses being stacked for an experiment in a commercial lens holder.}
    \label{fig:packaging}
\end{figure}

We fabricate our support disks from copper alloy (bronze) with diameters within a tolerance of 2-\SI{3}{\micro\meter}. Prior to laser ablation, the diamond plate in the form of a truncated cone is pressed carefully into the disk to avoid any tilt. Once the diamond is properly set, the lens profile can be micromachined in the centre of the disk. To ensure that both the back- and front- paraboloidal sections overlap, we mount the lens in a small stationary v-block shown in Fig.~\ref{fig:packaging}(a). We also use fiducial markings on the coin to align them to the v-block to determine the lens orientation during ablation - see Fig.~\ref{fig:packaging}(b). By packing our diamond lens in the same form factor as the industry-standard they can be readily integrated in beamlines using existing hardware (e.g. pin-holes, spacers, lens cases, v-blocks and transfocators) coexisting harmoniously with already acquired lenses. Figure~\ref{fig:packaging}(c) shows lens stacks in a commercial lens holder.

\subsection{Visible-light metrology}\label{sec:vismet}

\begin{figure}
    \centering
    \includegraphics[width=0.8\columnwidth]{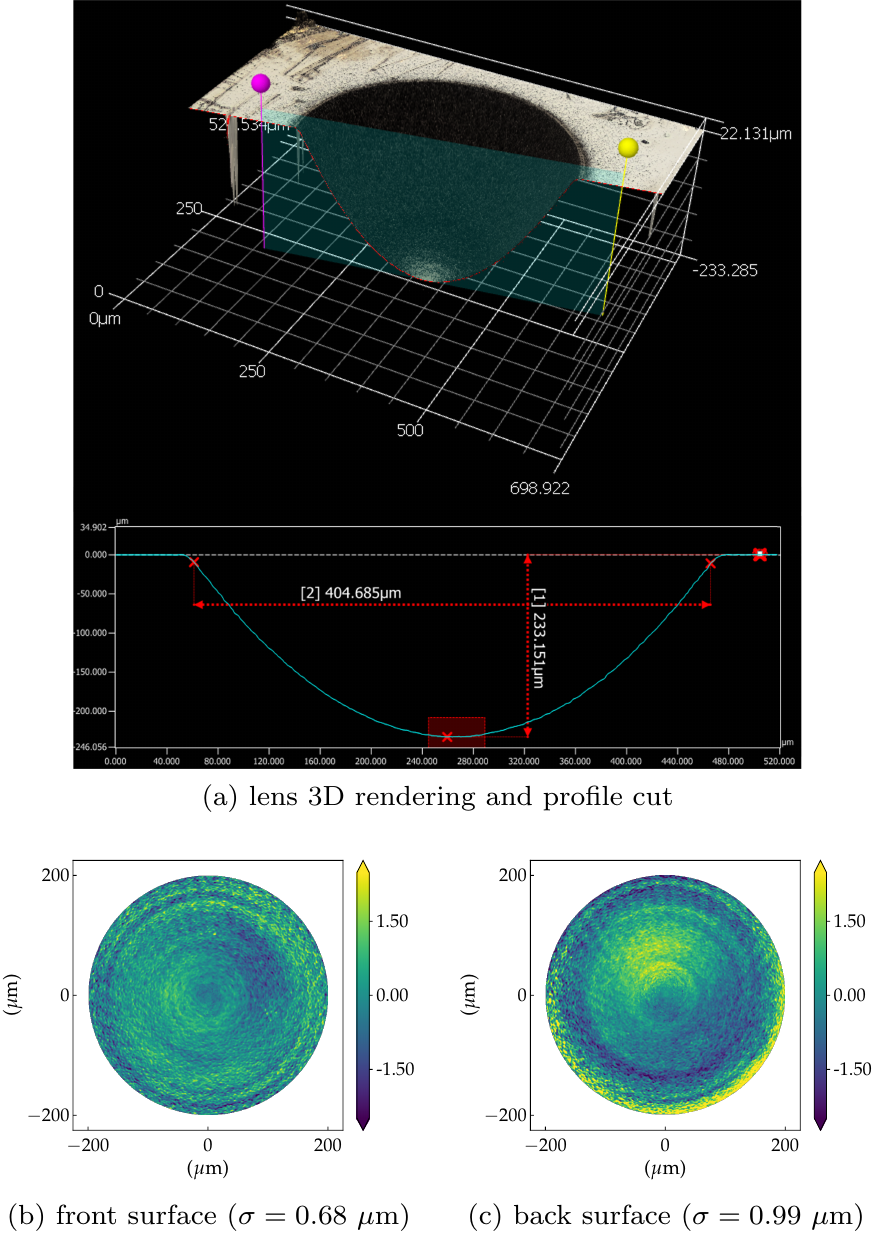}\vspace{-0.6cm}
    \caption{Laser scanning confocal microscopy of a unpolished diamond lens. (top) 3D reconstruction of front refractive surface. (bottom) residual profile after extraction of a paraboloid fit.}
    \label{fig:confocal_shape_error}
\end{figure}

Despite not being implemented as an in-situ measurement, we use scanning confocal laser microscopy for fine-tuning the ablation and post-polishing processes. This preliminary inspection tool is able to quickly provide information regarding the geometric aperture, refracting surface penetration depth, parabolic shape and radius of curvature of individual refracting surfaces as shown in Fig.~\ref{fig:confocal_shape_error}(a). 

Removing the best paraboloid fit from the metrology data, we are able to determine the figure errors as displayed in Fig.~\ref{fig:confocal_shape_error}(b) and (c), showing we achieve sub-micron rms figure errors for the laser-machined paraboloid. One of the limitations of this technique is the accurate determination of the mutual alignment of the measurements from the front- and back- surfaces in a bi-concave lens. Also, while measuring as-ablated surfaces is relative straightforward due to the elevated roughness and light scattering, polished lenses with their smooth surfaces and transparency are very difficult to measure using visible light confocal microscopy. A workaround to this inconvenience is applying a very thin conformal coating to reduce the transparency to visible light and make the laser scanning measurement possible.

\section{Lenses and lens stack measured in this work}

In this paper, we show experimental results obtained using individual lenses of three different types as well as results of three different lens stacks. The three types of individual lenses are a) unpolished $R=\SI{100}{\micro\meter}$ diamond lenses; b) polished $R=\SI{100}{\micro\meter}$ diamond lenses; and c) commercial $R=\SI{50}{\micro\meter}$ Be lenses produced by RXOptics \cite{lengeler_company_2021} - where $R$ represents the apex radius of curvature of the sagittal and meridional parabolic sections. The three different lens stacks are: \textit{A}) a stack of $N$=10 of the  unpolished $R=\SI{100}{\micro\meter}$ diamond lenses; \textit{B}) a stack of $N$=10 of the  polished $R=\SI{100}{\micro\meter}$ diamond lenses; and \textit{C}) a stack of $N$=11 of the commercial $R=\SI{50}{\micro\meter}$ Be lenses. The diamond lenses have a crystal thickness $L_\textrm{C*}$ of roughly \SI{500}{\micro\meter} (the lens thickness varied between \SI{471}{\micro\meter} and \SI{515}{\micro\meter}) and an a distance between the apices of the paraboloidal sections ($t$ or web thickness) of about $\sim\SI{20}{\micro\meter}$. The Be lenses are pressed into $L_\textrm{Be}=$1\,mm thick Be discs and have an apex width of about \SI{30}{\micro\meter}. 

At first glance, it seems problematic to compare $R=\SI{100}{\micro\meter}$ diamond lenses with $R=\SI{50}{\micro\meter}$ Be lenses. However, this comparison is actually very reasonable. First, the focusing strength of a single lens can be characterised by its focal length $f=\frac{R}{2\delta N}$ with $N$=1 for a single lens, $R$ the lens radius, and $\delta$ the refractive index decrement of the used element. At photon energies between 5\,keV and 50\,keV, the ratio $\delta_{C*}/\delta_{Be}$ varies little at around 2.14, see Appendix \ref{appendix:delta}. This makes a $R=\SI{100}{\micro\meter}$ diamond lens about equivalent to a $R=\SI{50}{\micro\meter}$ Be lens. Secondly, as the physical aperture of a lens is given as  $A_\textrm{phy}=2\sqrt{(L-t)\cdot R}$, and in our case we have $L_\textrm{Be}R_\textrm{Be}\approx L_\textrm{C*}R_\textrm{C*}$, we obtain also very similar physical apertures of $R\approx\SI{440}{\micro\meter}$.

Our two diamond lens stacks contain $N$=10 individual lenses (unpolished or polished). We can fine tune the number $N_\textrm{Be}$ of Be lenses to get the best equivalent Be lens stack as follows:
\begin{equation}\label{eq:Nbe}
   N_{Be}= N_{C^*} \frac{\delta_{C^*}}{\delta_{Be}}\frac{R_{Be}}{R_{C^*}}\approx  10.7
\end{equation}
We thus use $N$=11 lenses for the Be lens stack.


\section{At-wavelength metrology via x-ray speckle tracking}\label{sec:xsvt}

Although helpful in order to study surface roughness and measure the depth and aperture of the machined lens, the initial inspection using visible-light metrology after lens production suffers from limitations:  it only probes one side of a bi-concave lens at a time and it is insensitive to sub-surface defects; it also does not measure the alignment and mutual tilt of the front- and back-focusing surfaces. For the polished lenses, little signal returns from the steeper parts of the parabolic surfaces. To overcome those drawbacks and recover figure errors in projection approximation, we use x-ray (near-field) speckle vector tracking (XSVT) \cite{berujon_x-ray_2020} which is an at-wavelength metrology technique.

In the differential metrology mode of XSVT, the experiment consists of a monochromatic beam with sufficient lateral coherence illuminating a random static modulator and projecting the speckle pattern generated by it onto a 2D imaging detector. A reference dataset is taken by transversely shifting the random modulator across the beam and registering N images. A lens (-stack) is then introduced in the beam downstream of the speckle-modulator at a distance \textit{d} from the detector. A second data-set is taken by once again scanning the random modulator across the beam and registering N images at the exact positions where the reference dataset was taken. By comparing both datasets, the lateral displacement map $\nu_x(x,y),\nu_y(x,y)$ of the speckle pattern at the detector plane is calculated. With the knowledge of the distance \textit{d}, the deflection angle $(\alpha_x,\alpha_y)\approx(\nu_x,\nu_y)/d$ is retrieved. This deflection angle is related to the beam phase gradient $\nabla\phi$ by the wavenumber $k$. By numerical integration of the phase gradients obtained experimentally, the beam phase $\phi(x,y)$ is obtained. The projected thickness of the probe is calculated as $\Delta_z(x,y)=-\phi(x,y)/k\delta$. The figure errors in projection approximation are recovered by removing a paraboloid of revolution (2D focusing). 
The XSVT technique is described in more detail in \S2.2.3 \cite{berujon_x-ray_2020}
and \S3.1.1 from \cite{berujon_x-ray_2020-1}. 

Our XSVT experiments were performed at the ESRF-EBS beamline BM05 \cite{ziegler_esrf_2004} using a monochromatic beam - Si(111) double crystal monochromator - at 17~keV for single-lens metrology (\S\ref{sec:single_lens}) and 30~keV for the stack measurements (\S\ref{sec:lens_stack}). The change in energy for the stack measurement is to keep \textit{d} sufficiently large to avoid the speckle grains collapsing into each other, causing the tracking algorithm to fail. The measurements at lower energies used stacked cellulose acetate membrane filters with pore size $\sim\SI{1.2}{\micro\meter}$ as the speckle generator, while the measurements at 30~keV used stacked sheets of  1200 grit silicon carbide abrasive paper. The random modulators were mounted on piezoelectric nano-positioners to ensure high position repeatability between both data-sets. The distance \textit{d} between probe and detector was kept at 800~mm for the 17~keV measurements and 500~mm for the stacked lenses metrology. The detector was a pco.edge sCMOS sensor coupled to a 10$\times$ microscope objective imaging a $\sim\SI{17}{\micro\meter}$ thick GGG:Eu scintillator. The effective pixel size is $\SI{0.635}{\micro\meter}$. The lateral resolution has been evaluated at 17\,keV only, where it is better than \SI{1.5}{\micro\meter}.

This section presents the XSVT results of unpolished- and polished x-ray lenses and compares their shape errors to those of the equivalent $R=\SI{50}{\micro\meter}$ Be lenses of similar geometric aperture. A selection of these lenses is then used to form three stacks of similar focusing strength that are also measured and presented: $10\times$ unpolished, $10\times$ polished diamond and $11\times$ Be lenses. 

\subsection{Individual lens measurements}\label{sec:single_lens}

The metrology of individual lenses allows quantification of the figure errors of the lenses after the laser ablation machining process and can  help to fine tune this process. Furthermore, it can investigate the effect of the post-polishing process upon the figure errors of the x-ray lenses.

\begin{figure}
    \centering
    \includegraphics[width=0.7\columnwidth]{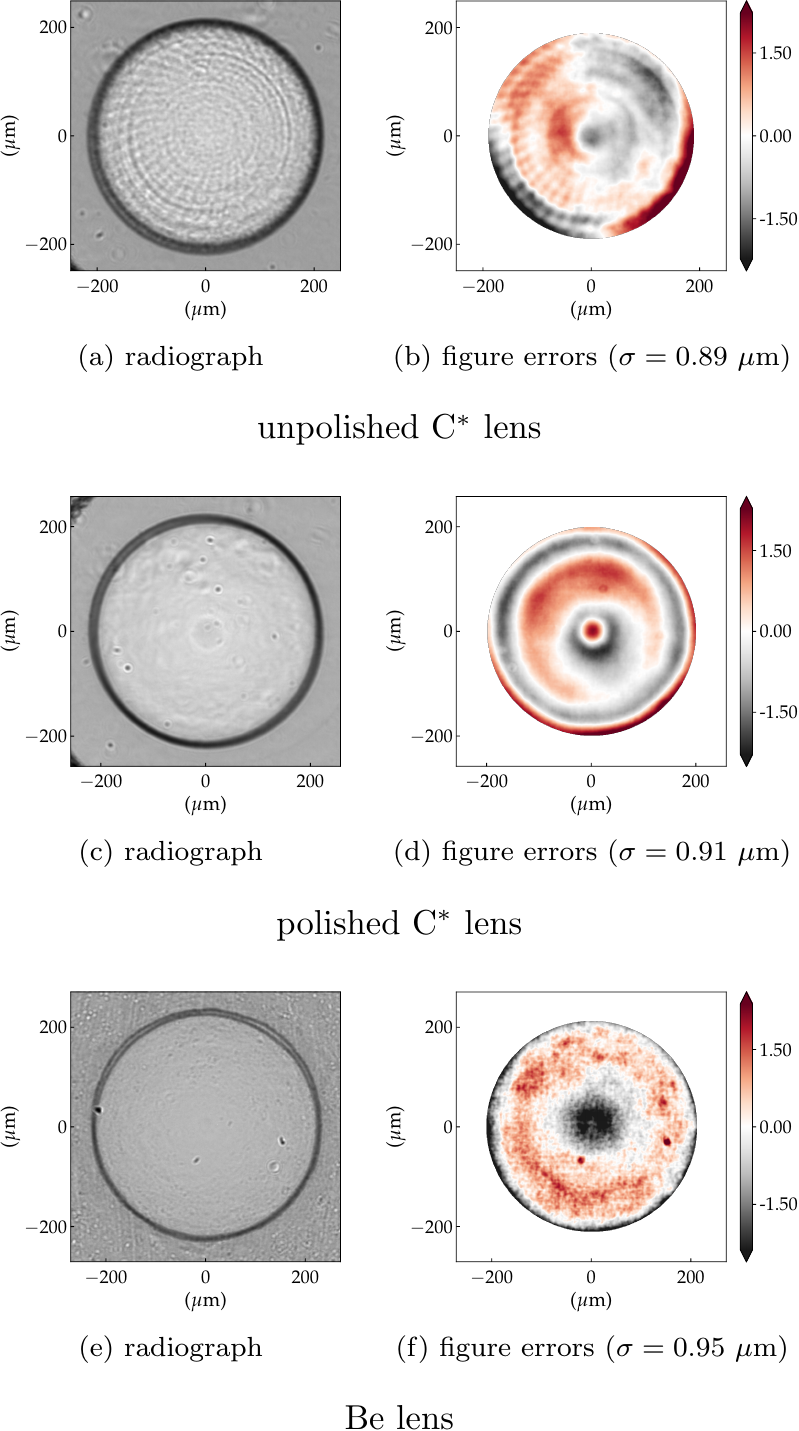}\vspace{-0.6cm}
    \caption{Single lens radiograph and figure error for each investigated lens type: unpolished $R=\SI{100}{\micro\meter}$ C$^*$ lens (top,), polished C$^*$ lens (middle) and $R=\SI{50}{\micro\meter}$ Be lens (bottom). Radiographs were taken 800~mm downstream of the sample, enhancing edge effects in phase-contrast imaging (dark rings delimiting the lens geometric aperture).}
    \label{fig:single_overview}
\end{figure}

Figure~\ref{fig:single_overview} shows representative examples of lens radiographs and figure errors for the three lens types used in this work. 
The presence of concentric circle-like- and bent-radial structures in the unpolished diamond lens radiography is evident - these features are also present in the accumulated figure errors for the same lens - cf. Fig.~\ref{fig:single_overview}(a)~and~(b). These features originate from the rough surface after the ablation process and are completely removed by  polishing  as shown in Fig.~\ref{fig:single_overview}(c)~and~(d). The post-polishing, however, introduces rotational symmetric figure errors to the lens; without significant change to the figure error rms value ($\sigma$) over the useful aperture. 

The Be lens radiograph is very homogeneous. The rms figure error of the Be lens is of similar size as that of both diamond lenses. However the figure errors of the Be lens can be well described by only lower order Zernike polynomials, i.e. they have a smoother long range order, plus a high frequency short range noise, possibly due to the  Be microstructure. 

The observations based on Fig.~\ref{fig:single_overview} are representative of all measured unpolished and polished diamond and Be lenses.

\begin{figure}
    \centering
    \includegraphics[width=\columnwidth]{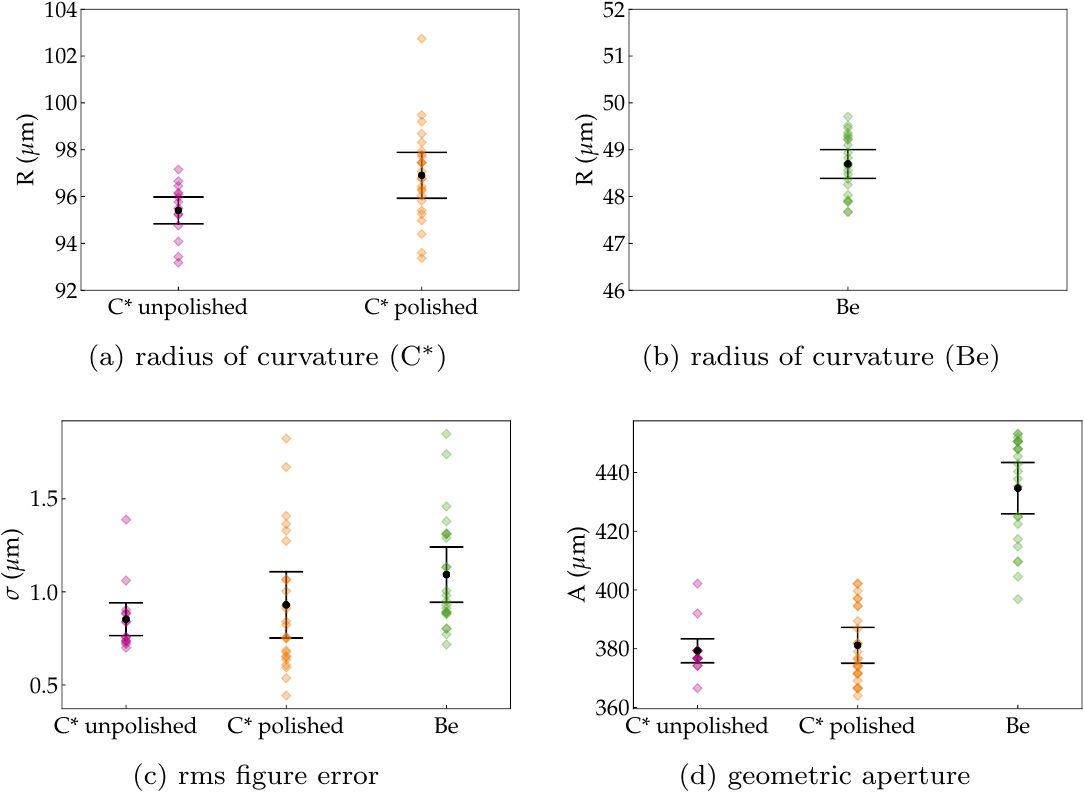}\vspace{-0.6cm}
    \caption{Dispersion plots of the lenses main figures of merit obtained with XSVT metrology of individual lenses.}
    \label{fig:single_stats}
\end{figure}

Fig.~\ref{fig:single_stats} summarises some key parameters (lens radius, rms figure error, and aperture) for 14 unpolished- and 25 polished diamond lenses comparing them against the metrology of 24 equivalent Be lenses. We can see that polishing the diamond lenses has the effect of slightly increasing the radius of curvature $R$. The unpolished lenses have a mean radius of curvature of $R=95.4\pm\SI{0.6}{\micro\meter}$, while the post-processed lenses have $R=97~\pm\SI{1}{\micro\meter}$ - Fig.~\ref{fig:single_stats}(a). As a reference, the Be lenses have  $R=48.7\pm\SI{0.3}{\micro\meter}$ - cf. Fig.~\ref{fig:single_stats}(b). The target radii are $\SI{100}{\micro\meter}$ for C$^*$ and \SI{50}{\micro\meter} for Be lenses. 

The figure errors of the diamond lenses (unpolished and polished) are slightly lower than those of commercial beryllium lenses: $\sigma=0.85\pm\SI{0.09}{\micro\meter}$, $\sigma=0.9\pm\SI{0.2}{\micro\meter}$ and $\sigma=1.1\pm\SI{0.1(5)}{\micro\meter}$ respectively. Although the mechanical polishing does slightly increase the nominal figure error and more significantly its dispersion, they are still close to the values measured on the commercial Be lenses - cf. Fig.~\ref{fig:single_stats}(c). Note that although the rms value of the figure errors is not significantly altered by the polishing, the spatial distribution is - see Fig.~\ref{fig:single_overview}. 

Lastly, the geometric aperture of the diamond lenses is systematically smaller than those of the equivalent Be lenses as shown in Fig.~\ref{fig:single_stats}(d). The probable reasons for this difference are threefold: \textit{a)} mismatch between the designed and executed penetration depth of the parabolic section (refracting surface) resulting in an increased distance between parabolic surfaces ($t$); \textit{b)} difference between real and expected diamond thickness $L$; \textit{c)} the radius of curvature $R$ of the lens. Mismatch between the penetration depths and misalignment between front and back refracting surfaces of the Be lenses explain the lower tail in Fig.~\ref{fig:single_stats}(d). These fabrication issues also occur in the production of the diamond lenses but are currently less recurrent. Another factor that contributes to the apparent reduction in the geometric aperture is that diamond gives stronger edge contrast in phase-contrast imaging when compared with beryllium at the same energy - this is manifested as dark ring delimiting the lens area in the radiographs from Fig.~\ref{fig:single_overview}. While it has no effect upon the geometric aperture, it  reduces the area from which metrology data can be extracted and consequently, the calculated  useful aperture plotted in Fig.~\ref{fig:single_stats}(d) is underestimated.

\begin{figure}
    \centering
    \includegraphics[width=\columnwidth]{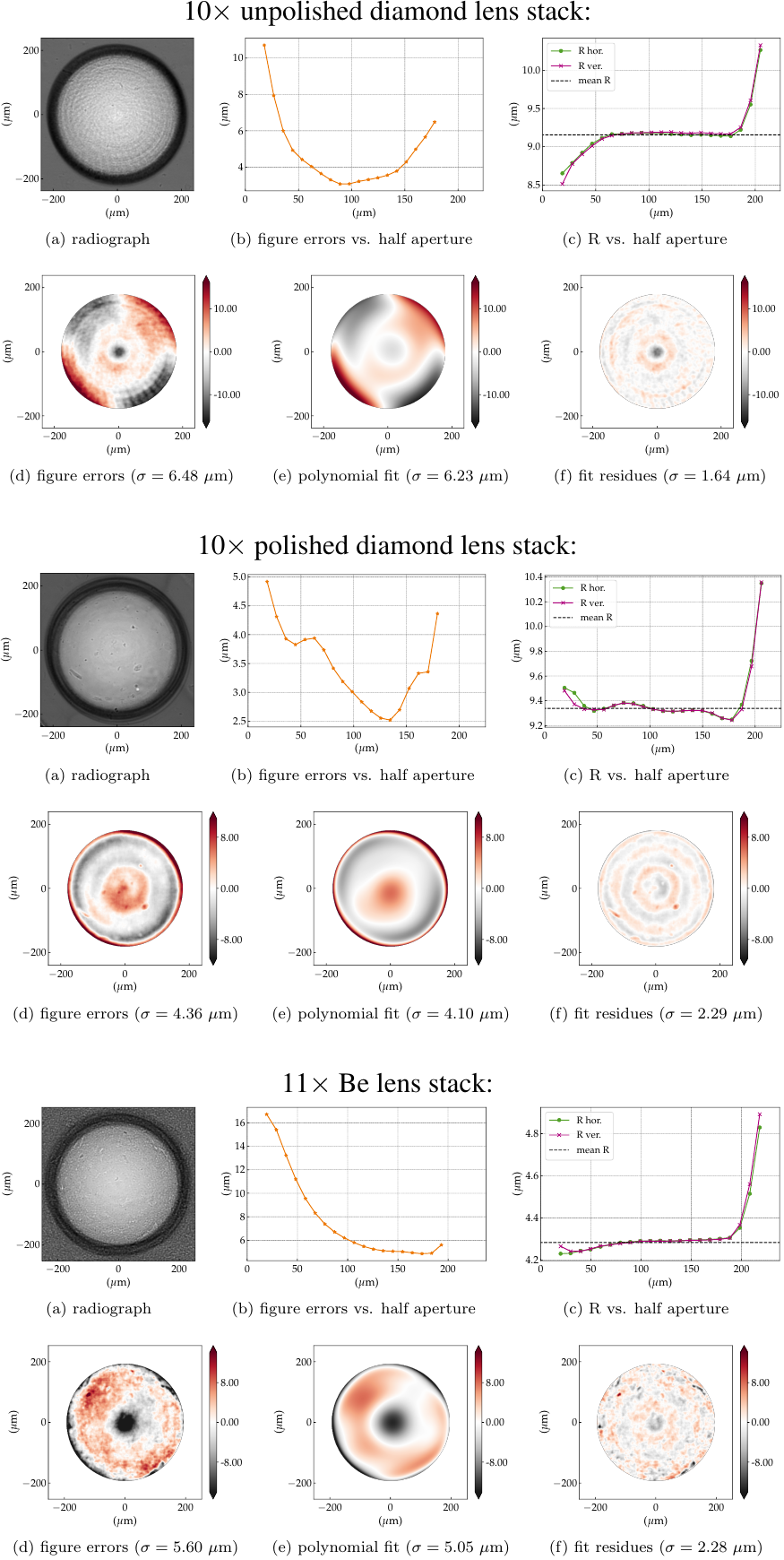}\vspace{-0.6cm}
    \caption{Lens stack metrology using XSVT. Top: $10\times$ unpolished diamond lenses, middle: $10\times$ polished diamond lenses, bottom: $11\times$ Be lenses. The coefficients of the polynomial decomposition in (e) are shown in Fig.~\ref{fig:Z_stack}.}
    \label{fig:stacks}
\end{figure}

\subsection{Measurements of lens stacks}\label{sec:lens_stack}

\begin{figure}
    \centering
    \includegraphics[width=\columnwidth]{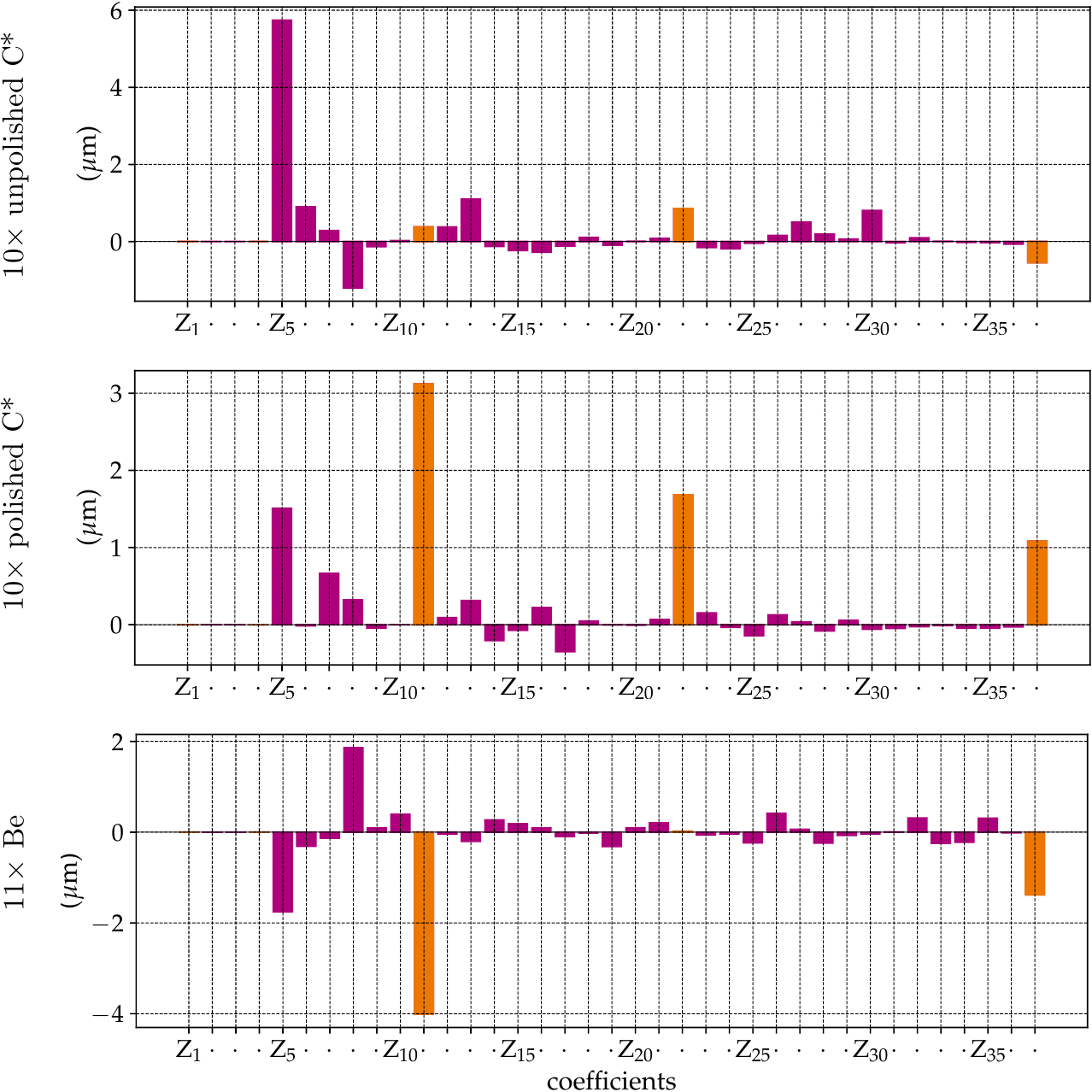}\vspace{-0.6cm}
    \caption{Zernike circle polynomial decomposition of the error profiles in Fig.~\ref{fig:stacks}. The terms $Z_1$ to $Z_4$ are suppressed as they account for piston, $x$ \& $y$ tilts and defocus, respectively and are not strictly optical aberrations. The terms $Z_5$ and $Z_6$ represent astigmatism, $Z_7$ and $Z_8$ show coma, $Z_9$ and $Z_{10}$ show tetrafoil aberrations. $Z_{11}$ stands for spherical aberration. $Z_{12}$ onward are higher order variations of the aberrations terms from $Z_{5}$ and $Z_{11}$. The orange bars are rotationally symmetric indicating primary to tertiary spherical aberrations.}
    \label{fig:Z_stack}
\end{figure}

X-ray lenses are generally used as CRL stacks. The metrology of stacked lenses is important because a) it allows simulation of the performance of the lens stack \cite{celestre_modelling_2020} and b) it permits the development of strategies for the mitigation of aberrations - see \S\ref{sec:corrector}. The measurements of 10 unpolished-, 10 polished diamond and 11 Be lens stacks are shown in Figs.~\ref{fig:stacks}~and~\ref{fig:Z_stack}. 

The characteristic laser machining marks on the rough surface of the unpolished diamond lenses, i.e. circle-like- and bent-radial structures, are still present when the lenses are stacked - see Fig.~\ref{fig:stacks}~(top row - a and d). For this particular set of lenses, the asymmetric figure errors tend to reinforce due to unfavourable azimuthal lens alignment inside the lens cassette - Fig.~\ref{fig:Z_stack}~(top row) - which could be mitigated by rotating individually the lenses around the optical axis \cite{osterhoff_aberrations_2017}. 

The polished diamond lens stack shows a more homogeneous radiograph, and a complete change in the topography of the error distribution - as compared to the unpolished stack - which displays stronger rotational symmetry, Fig.~\ref{fig:stacks}~(middle row - d) and orange bars in Fig.~\ref{fig:Z_stack}~(middle row). The change in error distribution - already visible in the metrology of individual lenses - is expected from the post-polishing process presented in \S\ref{sec:polishing}. 

The spherical aberrations of the diamond lenses, here accentuated by the post-polishing, are also commonly found in commercial Be lenses as presented in Figs.~\ref{fig:stacks}~and~\ref{fig:Z_stack}~(bottom row) and extensively reported \cite{celestre_modelling_2020, dhamgaye_correction_2020, seiboth_hard_2020}. The reason here is not a polishing process, but imperfections in the plastic forming process used for the lens manufacture or in the manufacturing of the paraboloidal punches on a lathe. A more quantitative analysis of the accumulated figure errors show that the polished lens stack has rms figure errors of $\sigma=\SI{4.36}{\micro\meter}$ over the entire useful aperture against $\sigma=\SI{5.60}{\micro\meter}$ of the Be lens stack. Although in terms of figure errors polished diamond lens stacks compare favourably to Be stacks, the optical path difference ($\textrm{OPD}=-k\delta\sigma$) still compares unfavourably for diamond lenses due to their higher index of refraction. The dependence of the rms figure errors upon the lens half aperture is shown in Fig.~\ref{fig:stacks}(d) for all three stacks.

\subsection{Possible correction of figure errors}\label{sec:corrector}

\begin{figure}
    \centering
    \includegraphics[width=\columnwidth]{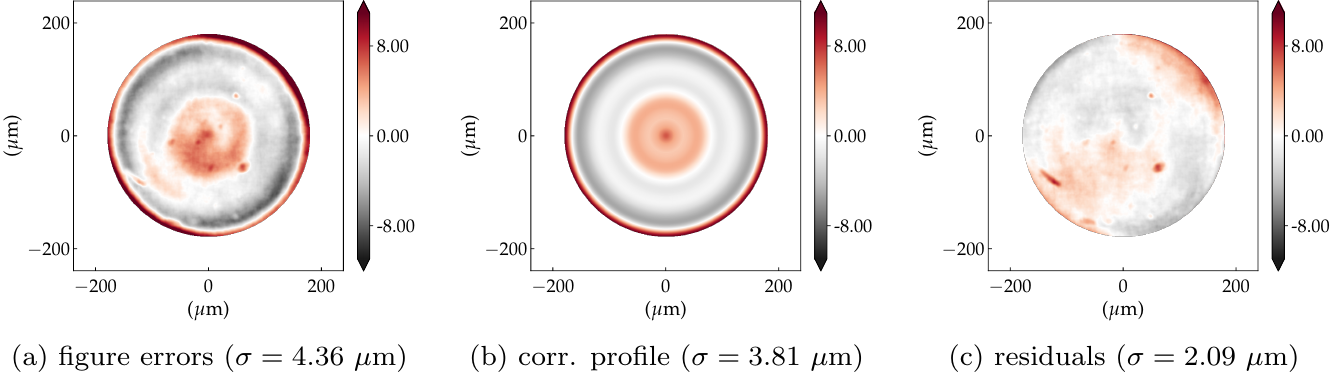}\vspace{-0.6cm}
    \caption{a) the measured figure errors of the 10$\times$ polished diamond lens stack. b) the closest calculated azimuthally symmetric figure error approximation of a). A phase corrector of inverted thickness variation profile is required to reduce the wavefront aberrations. c) the calculated residual effective figure errors expected from the combination of the lens stack and phase corrector.}
    \label{fig:correctable}
\end{figure}

The accumulated profile error of the polished lens stack shown in Fig.~\ref{fig:stacks}~(middle row - d) has strong rotational symmetry, as shown by the amplitude of the orange bars (primary to tertiary spherical aberration) in Fig.~\ref{fig:Z_stack}~(middle). This is particularly amenable to reduction of the wavefront aberrations via the use of azimuthally symmetric refractive phase plates \cite{seiboth_hard_2020,dhamgaye_correction_2020}. Fig.~\ref{fig:correctable} shows the approximately 50\% reduction of the effective figure errors of the polished diamond lens stack which should be achievable by the implementation of an ideal azimuthally symmetric diamond phase-corrector calculated to minimise the spherical aberrations.  

\section{Beam caustics and 2D intensity profile cuts along the beam path}\label{sec:caustics}

Metrology of single lens elements and lens stacks is important, however on a synchrotron beamline, the aim of using focusing lenses is usually to achieve a small focused probe with minimal scattering, halo or side lobes. Using a 2D detector with high spatial resolution, we recorded the transverse x-ray beam profile in the vicinity of the focal plane, when using each of the lens stacks A, B and C.

The experiment was performed on ID06 \cite{kutsal_esrf_2019} using a 10\,keV beam after a Si(111) double crystal monochromator. Each stack was mounted on a motorised hexapod alignment stage positioned  54\,m from the x-ray source point ($\lambda_u=27$\,mm permanent magnet undulator), with identical experimental conditions (slit settings, vacuum pipes and in-air sections, etc.). A 2D CCD detector (Atmel TH7899M, $14\times\SI{14}{\micro\meter}^2 $ pixel size) captured the visible light generated by a \SI{25}{\micro\meter}  thick LuAG:Ce scintillator via a 10$\times$ objective lens, a 0.9$\times$ extension tube, and a 2.5$\times$ eye piece. The effective pixel size was \SI{0.62}{\micro\meter}  \cite{kutsal_esrf_2019}. The distance between the lens stack and the detector scintillator could be varied. At 10\,keV, the expected focal lengths of the lens stacks are given in Table~\ref{table:stacks}.

\begin{table}[]
    \caption{Characteristics of the lens stacks at 10\,keV and the consequent focusing parameters. $f$ is the calculated focal length of the stack, $p$ is the distance between the lens stack and the photon source and, $q$ is the image distance calculated from $f$ and $p$ via the thin-lens equation. $M$ is the magnification when imaging the source.  $L$ is the thickness of the lens material at the lens rim. $D_\textrm{phys}$ is the physical aperture. $D_\textrm{eff}$ is the effective aperture reduced by x-ray absorption in the thicker lens regions  - equation 27 of \cite{kohn_effective_2017} - for 10\,keV.}
    \centering
    \begin{tabular}{lcc|cccc|ccc}
          & $N$ & $R$ & $f$ & $p$ & $q$ & $M$ & $L$ & $D_\textrm{phys}$ & $D_\textrm{eff}$ \\
          & & [\SI{}{\micro\meter} ] & [mm] & [m] & [mm] & & [mm] & [\SI{}{\micro\meter}] & [\SI{}{\micro\meter}] \\ 
          \vspace{-0.4cm}
         \hline
          & & & & & & & & & \\
         C$^*$ & 10 & 100    & 685 & 54 & 694 & 78 & 0.5 & 436 & 231\\
         Be    & 11 & \ \ 50 & 666 & 54 & 674 & 80 & 1   & 442 & 361\\
    \end{tabular}
    \label{table:stacks}
\end{table}

\begin{figure}
    \centering
    \includegraphics[width=\columnwidth]{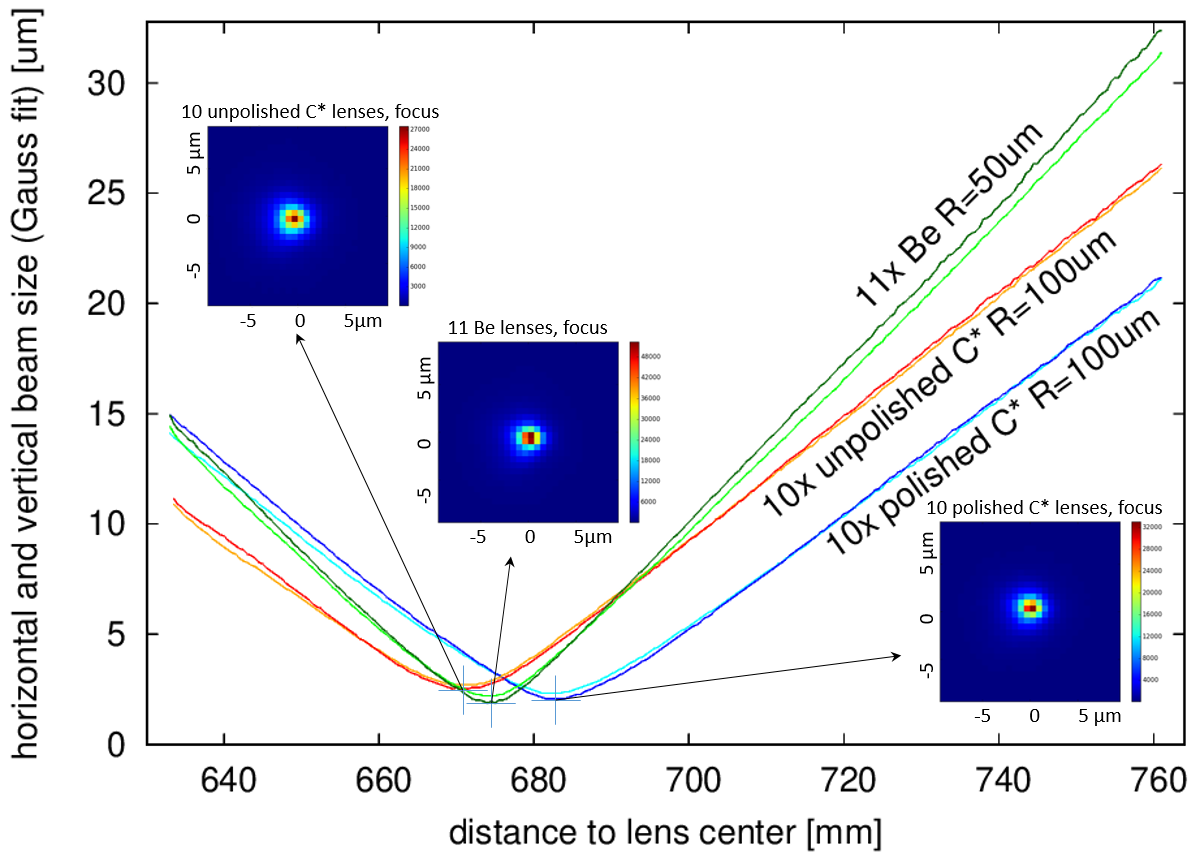}\vspace{-0.6cm}
    \caption{Beam caustics and profiles at the focal plane.}
    \label{fig:caustic}
\end{figure}

Fig. \ref{fig:caustic} shows the beam size (FWHM) as measured by the CCD camera and fitted with a 2D Gaussian while the detector is scanned along the x-ray beam path. All three lens stacks have a focal plane in the vicinity of 670\,mm and show similar minimum beam size, which is close to the optimal resolution of the CCD camera.

\begin{figure}
    \centering
    \includegraphics[width=\columnwidth]{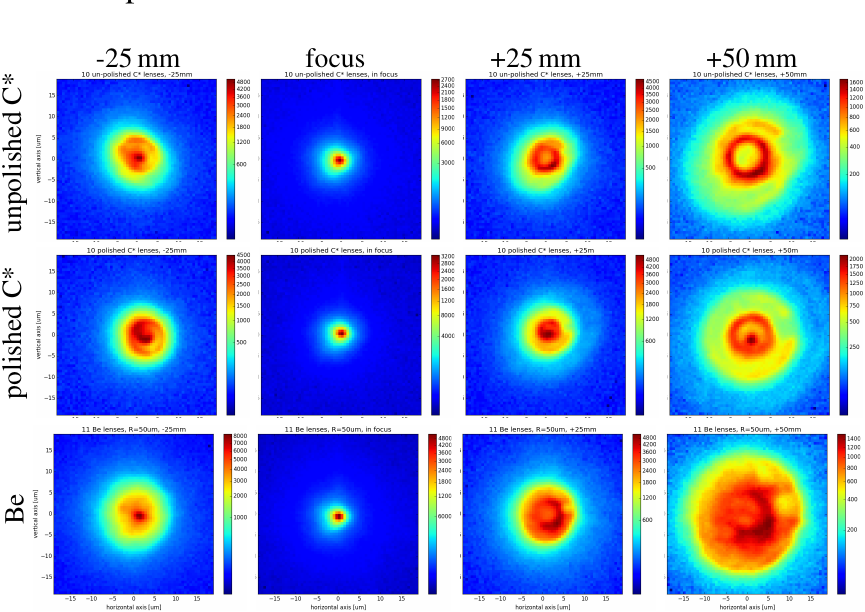}\vspace{-0.6cm}
    \caption{The x-ray beam intensity profile as measured 25\,mm upstream of focus, at focus, and 25\,mm and 50\,mm after the focus. Top row: 10 unpolished diamond lenses. Middle row: 10 polished diamond lenses. Bottom row: 11 beryllium lenses. Note that the colour range is different for each column. The colour range has been plotted using a power law with exponent 0.33 (gamma correction), in order to highlight less intense regions.}
    \label{fig:2Dimages}
\end{figure}

Fig. \ref{fig:2Dimages} presents images of the x-ray beam intensity distribution at fixed distances from the focal position, namely 25\,mm upstream of the focus, at focus, 25\,mm downstream and 50\,mm downstream of the focal position.

The beamsizes (both vertical and horizontal) for the Be lens stack increase faster when getting further from the focal plane, as compared to both diamond lens stacks. The origin for this behaviour is the larger effective aperture of the Be lens stack, see Table \ref{table:stacks}, which implies that more photons farther off the optical axis are transmitted by the Be lenses as compared to the diamond lenses. At low photon energies, 10\,keV for example, Be lenses are the more efficient focusing element, but this advantage is much reduced at higher x-ray energies. Here at 10\,keV, and with an incoming x-ray beam size upstream of the lenses of approximately 1$\,\times\,$1\,mm$^2$ that is slit down to $450\times\SI{450}{\micro\meter}^2$, the measured integrated detector signal after the Be lenses is about 1.8 times higher than in the case of the diamond lens stack. This is in agreement with a calculated transmission of a flat square beam of size  $450\times\SI{450}{\micro\meter}^2$ through a lens with physical lens aperture of \SI{440}{\micro\meter} and a web thickness $t_\textrm{min}$ of \SI{30}{\micro\meter} as a function of the projected thickness $\Delta(x,y)$, that is $T\propto\int \exp\big[-\mu \cdot \Delta (x,y)\big] \text{d}x \text{d}y$. For the beryllium and both diamond and lenses: $T^\textrm{on apert.}_\textrm{11\,Be}$=0.791, $T^\textrm{on apert.}_\textrm{10\,C*}$=0.463, giving a ratio of 1.71. As an example for higher energies, at 30\,keV, while maintaining the focal length in using 9 times as many lenses ($N_\textrm{Be}$=99, N$_\textrm{C*}$=90), the advantage in transmission of the Be stack is reduced to 1.27, via a similar calculation.   

\section{Wire scan measurements: beam size at focus} \label{sec:wire}

In order to overcome the lateral resolution limit of the 2D CCD camera system, we installed a $\diameter\SI{200}{\micro\meter}$  tungsten wire on precision y-z-translation stages (Newport MFA-PP), followed by a large surface area Si p-i-n diode connected to a pico-ampere meter. The tungsten wire was scanned both vertically and horizontally through the beam (stepsize of \SI{0.25}{\micro\meter}), and this experiment was repeated with slightly varying lens to wire distance. The measured intensity profile with the diode shows an S-shaped curve between a position outside the beam and when blocking the beam. After Gaussian fit of the numerically differentiated signal, we calculate the projected FWHM beam size in the horizontal and in the vertical directions. 

\begin{figure}
    \centering
    \includegraphics[width=\columnwidth,trim=0 0 0 12,clip]{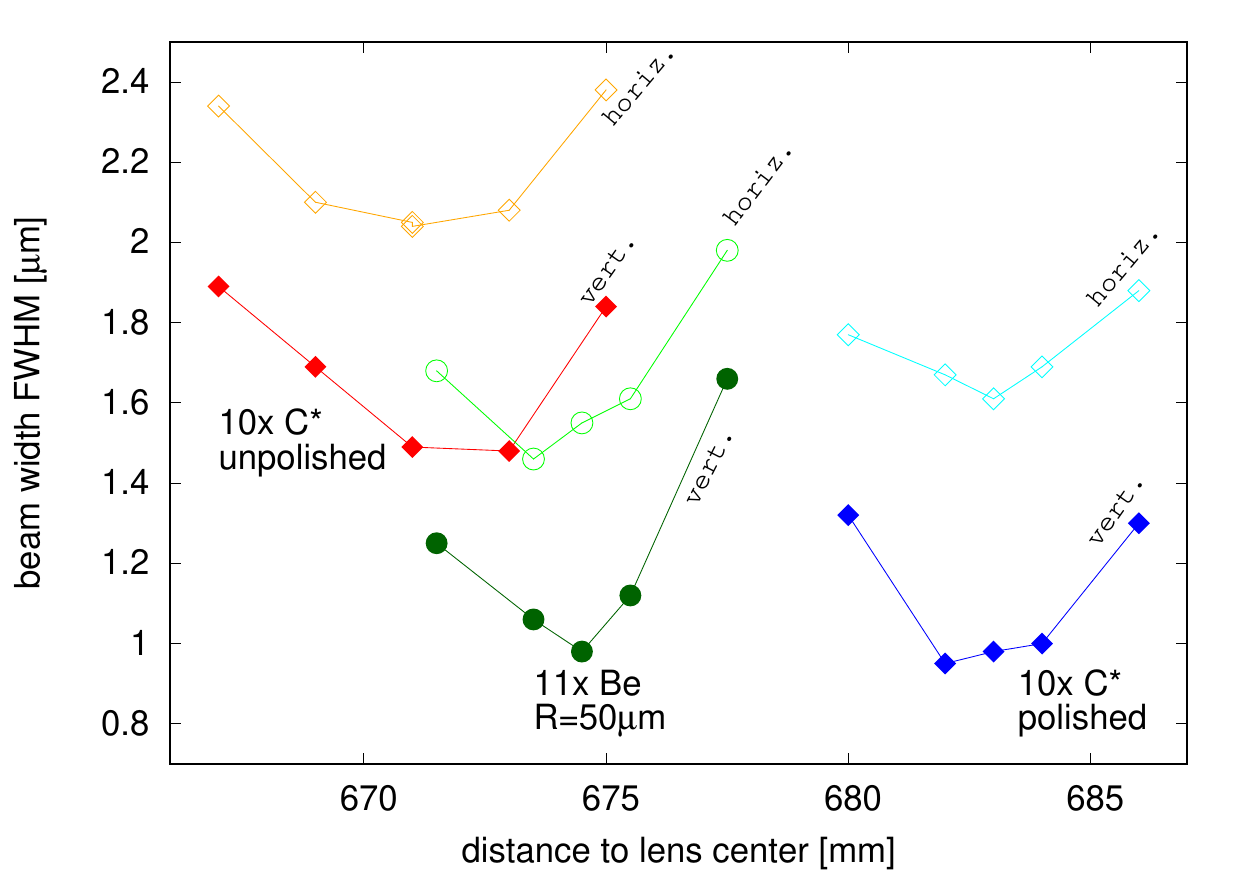}\vspace{-0.6cm}
    \caption{Beam sizes in the vicinity of the focal plane as measured by scanning a $\diameter\SI{200}{\micro\meter}$ tungsten wire through the x-ray beam.}
    \label{fig:wireOverview}
\end{figure}

\begin{figure}
    \centering
    \includegraphics[width=\columnwidth,trim=0 14 0 20,clip]{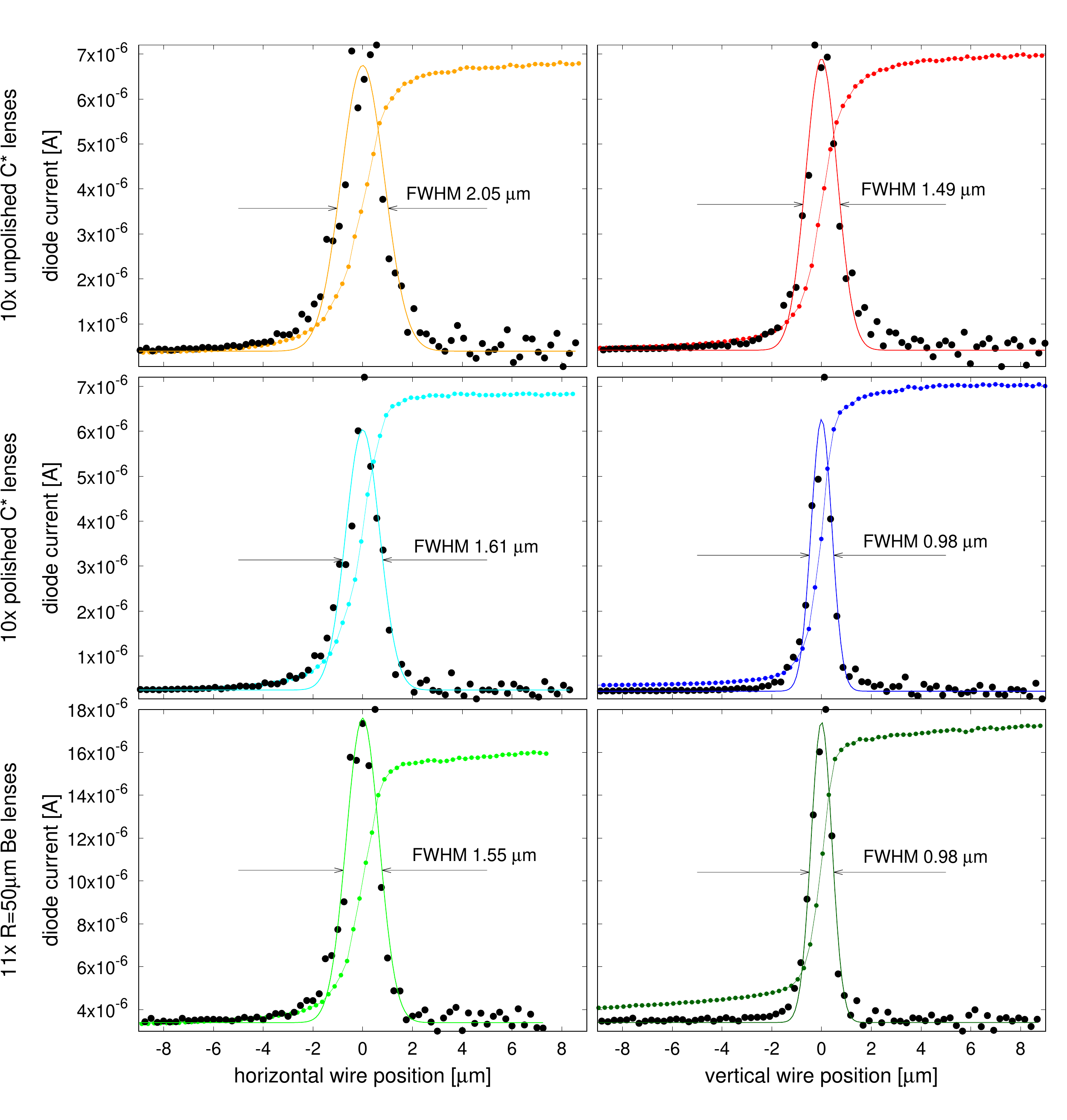}\vspace{-0.6cm}
    \caption{Smallest measured beam size using the wire scan technique for all three lens stacks (top: unpolished C* lenses, middle polished C* lenses, bottom Be lenses) We first take the numerical derivative $\Delta I_\textrm{diode}/\Delta x$ of the beam intensity measured via the current generated in a p-i-n photo diode (raw signal: coloured lines with points). This derivative (black dots) is fitted by a Gauss function (solid lines). Left column: horizontal beam size; right column: vertical beam size. }
    \label{fig:wireDeriv}
\end{figure}

Fig.\,\ref{fig:wireOverview} shows the measured FWHM beam size along both the vertical (dark colours, filled symbols) and the horizontal (bright colours, open symbols) direction - for the three lens stacks as a function of the distance center-of-lens-stack to wire. This figure is consistent with Fig.\ref{fig:caustic}, however the resulting FWHM are slightly lower, as the result is not broadened by the detector lateral resolution.

As can be seen from the position of the focal plane in Fig.\,\ref{fig:wireOverview}, the radii of the polished lenses are about 1.6\,\% larger than the radii of the unpolished ones. Polishing is removing material and thus tends to increase the lens radius. This is in agreement with the at-wavelength metrology results, see Fig.~\ref{fig:single_stats}(a). For the central position for each of the lenses in Fig.\,\ref{fig:wireOverview}, the corresponding raw data of the wire scan, its derivative (arbitrary units, not shown), and a Gaussian fit to this derivative are shown in Fig.\,\ref{fig:wireDeriv}.

The measured beam sizes are larger than expected. At an ESRF beamline, after the EBS upgrade, the FWHM X-ray source size in the 16 bunch filling mode is $\SI{70}{\micro\meter}\times\SI{18}{\micro\meter}$ (horizontal $\times$ vertical). In geometric optics, using $M$ of $\approx\,79$ from Table \ref{table:stacks}, this would give a focused beam size of approximately $\SI{0.88}{\micro\meter}\times\SI{0.23}{\micro\meter}$. Considering Gaussian optics and the diffraction limit corresponding to the effective aperture of each lens stack, as well as slightly different focal lengths, that changes the expected FWHM beam size in the focal plane to  $\SI{0.883}{\micro\meter}\times\SI{0.255}{\micro\meter}$ for Be and $\SI{0.924}{\micro\meter}\times\SI{0.314}{\micro\meter}$ for diamond, where we followed Eq.\,44 of \cite{lengeler_imaging_1999}. The vertical values are diffraction limited and differ significantly for the two materials, as a consequence of the different effective apertures, see Table \ref{table:stacks}. A point source would be imaged to a \SI{0.10}{\micro\meter} image for the case of the Be lens stack, and a \SI{0.20}{\micro\meter} image for the diamond lens stack. Our measurements for the polished diamond lens stack are larger by a factor 1.7 in the horizontal and a factor 3 in the vertical. For the Be lens stack, these factors are 1.6 (horizontal) and 3.7 (vertical).

Above calculations consider ideal lens shapes, however. As we measured larger values, these deviations can come from lens shape errors, lens surface roughness effects (as seen in the larger focal sizes of the unpolished lenses), and also possibly vibrational beam size broadening by a vertically deflecting, cryogenically cooled double crystal monochromator as present on ID06. The effect of the lens shape errors can be evaluated via simulations, as we have measured the error profiles for all lens stacks.


\section{Simulations of the 2D profile cuts along the beam path and beam size at focus} \label{appendix:srw}

We have used the metrology profiles obtained with XSVT in §\ref{sec:lens_stack} - see Fig.~\ref{fig:stacks}(d) - to simulate the x-ray beam focusing by the lens stacks used in this work. These partially-coherent simulations were performed using SRW's macro-electrons method \cite{chubar_accurate_1998, chubar_development_2011} and the refractive optics Python library described in \cite{celestre_recent_2020}, which implements the modelling of phase imperfections in refractive optics presented in \cite{celestre_modelling_2020}. The calculations here replicate the ID06 beamline in similar conditions to the experiments described in \S\ref{sec:caustics} and \S\ref{sec:wire}. 

\begin{figure}
    \centering
    \includegraphics[width=\columnwidth]{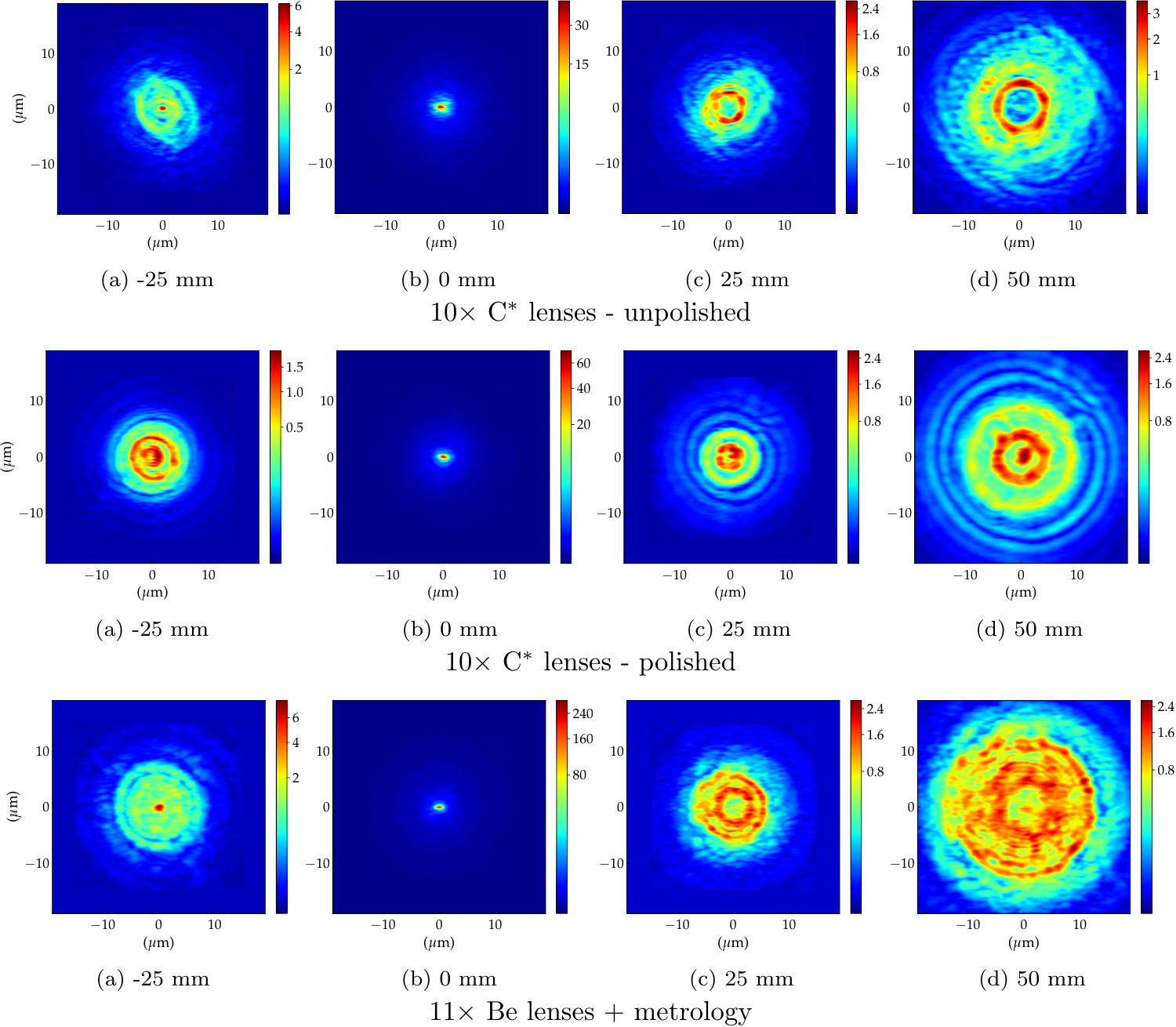}\vspace{-0.6cm}
    \caption{X-ray beam profile simulated 25\,mm upstream of focus, at the image plane, and 25\,mm and 50\,mm after the focal plane. Top row: 10$\times$ unpolished diamond lenses. Middle row: 10$\times$ polished diamond lenses. Bottom row: 11$\times$ beryllium lenses. Note the different intensity range for each column. The plots use a gamma correction ($\gamma=0.33$) in order to highlight weaker intensity regions.}
    \label{fig:srw_beam_cuts}
\end{figure}

The first row of Fig.~\ref{fig:srw_beam_cuts} shows the simulated focusing of the unpolished lens stack with strong similarity with the experimental data shown in the top row of Fig.~\ref{fig:2Dimages}. Features like the central lobe 25~mm upstream of the focal plane and the ring profile downstream are very well reproduced - see concentric quarter of a ring formation (in orange/yellow) at +50~mm in both experimental and and simulated data, for example. The tilted oval shapes at -25~mm and +25~mm and the orientation of the major axes are also reproduced. The simulations in Fig.~\ref{fig:srw_beam_cuts} - middle row - also reproduce important features from the polished stack, namely the concentric ring-like structures observed up- and downstream of the focal plane. The central lobe seen at +50~mm is also visible in the simulations, along with a few concentric rings -  also visible in Fig.~\ref{fig:2Dimages}. The beam focusing by the the Be lens stack shown in  Fig.~\ref{fig:srw_beam_cuts} (bottom row) is also in good agreement with the experimental data in Fig.~\ref{fig:2Dimages} - e.g. with a central lobe observed upstream of the focal plane and ring structures downstream. The simulations also describe well the beam scattering around the focal spot (0~mm) and the fact that the beam sizes at +50~mm are different. It is worth noting that the simulations have a far higher lateral resolution than the experimental data and do not suffer from any degradation (e.g. vibrations, beam instabilities, detector noise and etc.), hence the beam features are better defined and richer in details than the experimental data.

\begin{figure}
    \centering
    \includegraphics[width=0.9\columnwidth]{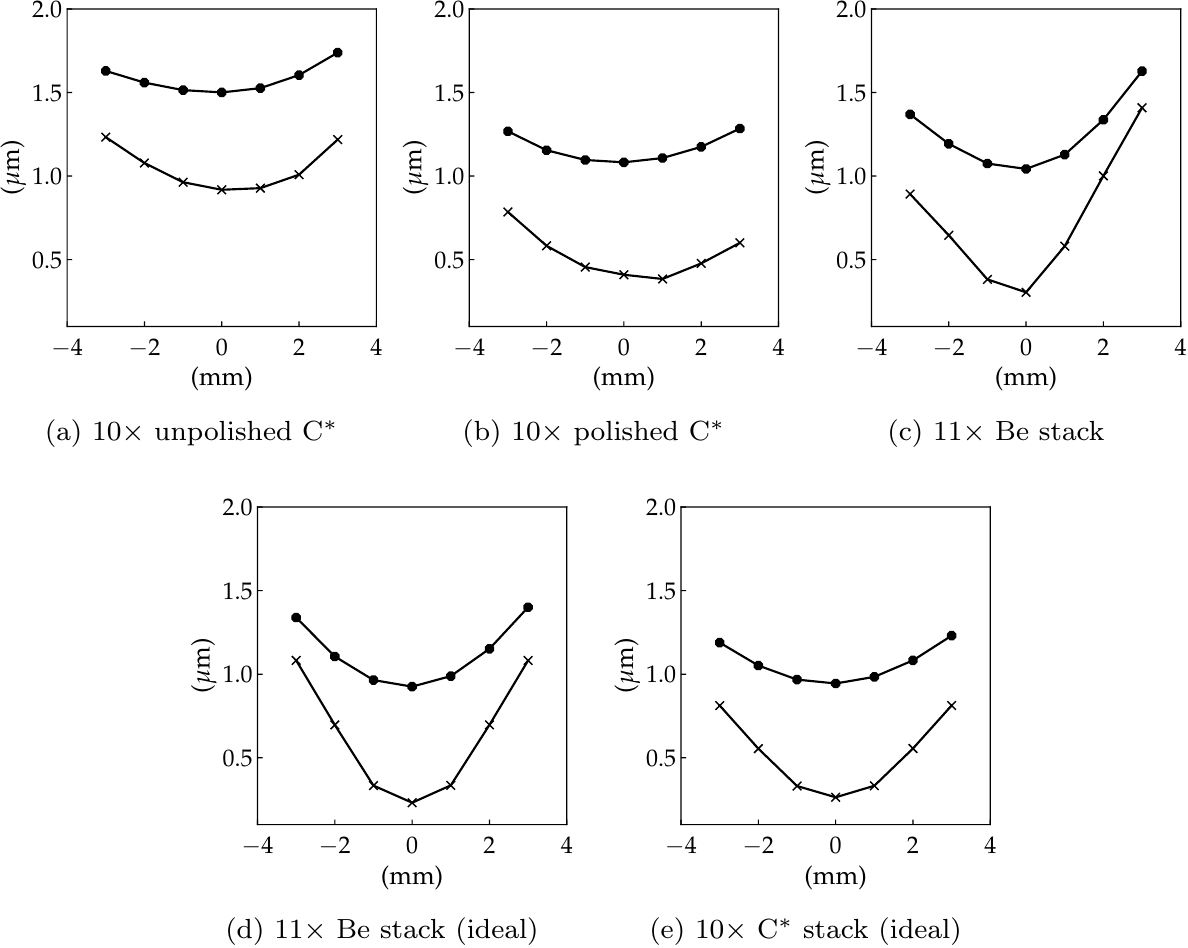}\vspace{-0.6cm}
    \caption{Beam sizes (FWHM) in the vicinity of the focal plane for different lens stacks. Horizontal values are represented by $-\bullet-$ and the vertical profile sizes by $-\times-$ markers.}
    \label{fig:srw_wire}
\end{figure}

The high resolution of the simulations can be used to evaluate the beam size around the focal plane. By projecting the intensity horizontally \& vertically and performing a Gaussian fit of the resulting profiles, we obtain results that can be compared with the measurements in \S\ref{sec:wire}. The simulated beam focusing can be seen on Fig.~\ref{fig:srw_wire}, which agrees qualitatively  with Fig.~\ref{fig:wireOverview}. The introduction of figure errors to the simulations show clear degradation of the focal spot size, which is more evident for the unpolished diamond lens stack. For reference, ideal focusing is presented in Fig.~\ref{fig:srw_wire}(d) and (e). In general, we see a worse performance for the unpolished C$^*$ stack and very similar performances for the polished C$^*$ and Be stacks, with the latter being slightly better. The steeper caustic from the Be stack (related to larger effective aperture and shorter focal length) is also shown by the simulations. Regarding experimental data (Fig.~\ref{fig:wireDeriv}), simulations are more optimistic in terms of beam sizes at the focal plane. These discrepancies, stronger for the vertical plane, might be partially explained by vibrations in the monochromator or thermal deformations of the first crystal of the monochromator.

A compilation of the beam sizes for the different stacks is presented in Table~\ref{tab:srw_sizes}.

\begin{table}[]
\label{tab:srw_sizes}
\caption{Summary of beam sizes in micrometers obtained from simulations and experimental data for different lens stacks.}
\centering
\resizebox{\textwidth}{!}{%
\begin{tabular}{rccccc|ccc}
\multirow{3}{*}{stacks:} & \multicolumn{5}{c|}{\textbf{diamond}} & \multicolumn{3}{c}{\textbf{Be}} \\ \cline{2-9} 
 & \multirow{2}{*}{ideal} & \multicolumn{2}{c|}{unpolished} & \multicolumn{2}{c|}{polished} & \multirow{2}{*}{ideal} & \multirow{2}{*}{simulation} & \multirow{2}{*}{experimental} \\ \cline{3-6}
 &  & sim. & \multicolumn{1}{c|}{exp.} & sim. & exp. &  &  &  \\ \cline{2-9} 
\multicolumn{1}{r|}{hor.:} & 0.94 & 1.50 & 2.05 & 1.08 & 1.61 & 0.93 & 1.04 & 1.55 \\
\multicolumn{1}{r|}{ver.:} & 0.26 & 0.92 & 1.49 & 0.38 & 0.98 & 0.23 & 0.30 & 0.98
\end{tabular}
}
\end{table}

\section{Small angle x-ray scattering}\label{sec:saxs}

\begin{figure}
    \centering
    \includegraphics[width=\columnwidth]{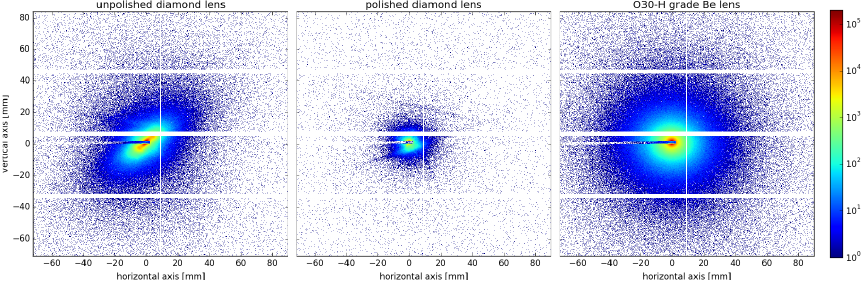}\vspace{-0.6cm}
    \caption{Raw 2D detector images showing the SAXS signal of an unpolished (left), a polished (middle) diamond and O30-H Be lens (right), taken at 8\,m lens to detector distance.}
    \label{fig:SAXSraw}
\end{figure}

The absorption of X-rays in a stack of lenses is an important characteristic of the lens for X-ray applications, since it determines the flux in the focused beam. Moreover, the lens material should generate the lowest intensity of background scattering. The intensity of the background signal observed in the tails of the focused beam \cite{gasilov_refraction_2017} is mainly caused by small-angle (x-ray) scattering \cite{guinier_diffraction_1939,glatter_small_1982}, which is primarily due to electron density fluctuations in the scattering volume (SAXS). The microstructure of the lens material can be either (i) single crystalline such as for the diamond lenses as presented here, (ii) polycrystalline (for Be lenses) \cite{roth_x-ray_2014}, or (iii) amorphous (e.g. glassy carbon lenses) \cite{artemiev_x-ray_2006}.  Additionally, the surface roughness at the air-lens interface also contributes to the total scattering background. It is therefore insightful to compare the SAXS intensities for the different lens materials with that originating from the surface finish. Here we present SAXS measurements for an unpolished and a polished diamond lens, and compared the total scattered intensity to a commercial Be lens manufactured from Materion O30-H grade material. SAXS measurements were carried out at ESRF-EBS beamline ID02 \cite{narayanan_multipurpose_2018} using an incident X-ray energy of 12.23\,keV, corresponding to a wavelength $\lambda$ = 1.013 \AA. In order to cover a wide q range, 0.002 $\leq q \leq$ 3 nm$^{-1}$ ($q$ is the magnitude of the scattering vector given by $q=4\pi/\lambda \sin{\theta/2}$, with $\theta$ the scattering angle), three different sample-detector distances were used: 31~m, 8~m and 1~m. The 2D scattering patterns were recorded using an Eiger2 4M detector with an active area of 155.1 x 162.2\,mm$^2$ and a pixel size of $75\times\SI{75}{\micro\meter}^2$. Figure\,\ref{fig:SAXSraw} shows the 2D SAXS pattern acquired at 8\,m sample-detector distance for the unpolished (left), the polished diamond (middle) and the reference O30-H Be lens (right). The SAXS was recorded with the X-ray beam centred on the thinnest part of each lens. The beam size on the sample was approximately $100\times\SI{100}{\micro\meter}^2$, with a photon flux of the order of 6.7$\times$10$^{11}$ photons/s. 

\begin{figure}
    \centering
    \includegraphics[trim= 5 12 18 12 , clip, width=0.8\columnwidth]{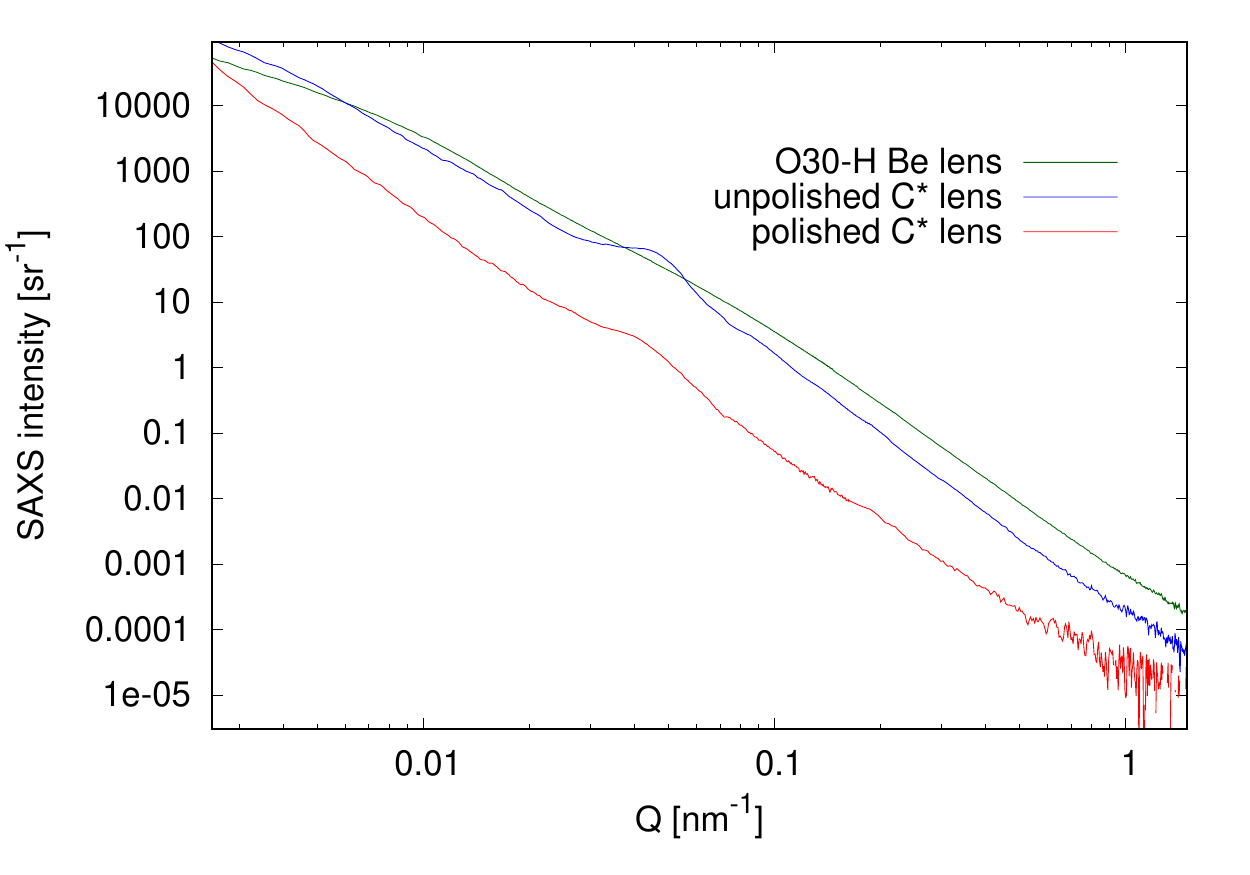}\vspace{-0.6cm}
    \caption{1D SAXS intensities for the 3 investigated lenses compared to the empty background. Note that the raw signal for the unpolished diamond lenses in Fig.\,\ref{fig:SAXSraw} exhibits a large anisotropy. In the azimuthally averaged data a correlation peak appears at $q=0.045$\,nm$^{-1}$. Due to the unknown thickness of the lenses the normalised intensity $I(q)$ is given in $\textrm{sr}^{-1}$.}
    \label{fig:IofQ}
\end{figure}

Figure\,\ref{fig:SAXSraw} shows that the polished diamond lens exhibits a lower background than the unpolished counterpart and also less than the Be reference lens. Additionally, the scattering of the unpolished lens appears highly anisotropic. As the scattering of the polished lens is isotropic, we relate the strong anisotropy to the surface condition before laser polishing. Possible explanations could be that a) the craters created by the laser ablation process are not radially symmetric, which might be the case if the focused laser beam is not of circular shape or that b) sub-micron scale surface texture is introduced by the linear polarisation of the laser beam as in laser-induced periodic surface structuring \cite{granados_photonic_2017}. Once this surface layer is removed (polished lens), the crater structure disappears and with it the asymmetric intensity distribution. Finally, the total scattering background of the polished diamond lens becomes low enough, so that the crystalline nature of diamond becomes visible represented by characteristic Kossel lines, see Fig\,\ref{fig:SAXSraw}(middle). Despite the observed anisotropy in the 2D SAXS of the unpolished diamond lens, we performed a full azimuthal integration. The obtained normalised 1D intensities, $I(q)$, are plotted in Fig.\,\ref{fig:IofQ}.
The lower SAXS signal after polishing the laser ablated diamond lenses is striking. Throughout the plotted $q$-range, the red curve is about two orders of magnitude lower than the curves of the unpolished diamond lens and of the O30H-grade beryllium lens. Note that there are beryllium grades which can yield lower SAXS signals (IF-1, IF-5 and IS50M), though this will improve the situation only by about one order of magnitude \cite{roth_x-ray_2014}. The correlation peak in the azimuthally averaged data at $q=0.045$\,nm$^{-1}$ of the unpolished lens is still visible in the polished lens, but much reduced. Via $2\pi/q$ we obtain a characteristic spacing of 140\,nm, which could be well explained by the aforementioned periodic structuring during the laser ablation process.

\section{Conclusion}

In this paper, we have presented single crystalline bi-concave 2D focusing diamond lenses that are comparable in quality to commercial Be lenses. These diamond lenses were produced via femto-second laser ablation followed by a mechanical polishing step. This reduces the surface roughness to an extent that the polished lenses yield a much reduced SAXS background. Focusing capabilities and lens figure errors are very close to those obtained by equivalent Be lenses. We have measured figure errors of the diamond lenses of \SI{1}{\micro\meter} rms over the full useful lens aperture of almost \SI{400}{\micro\meter}.

Whether single crystalline HPHT diamond is required or cheaper sc-CVD or polycrystalline grades can be used and give similar quality lenses will be investigated in future measurements. In that case, larger thickness crystals (e.g. 1\,mm thick as common for Be lenses) could allow an increase of the physical aperture or reduction of the lens radius while maintaining the current physical aperture. 

The diamond lenses presented here are mounted in common 12\,mm diameter lens frames and feature a reduced frame thickness of 1.2\,mm, which can be beneficial for high energy focusing applications as the total lens stack length can be reduced. 

As noted in \cite{serebrennikov_optical_2016}, diamond lenses are the material of choice for x-ray lenses at higher x-ray energies, i.e. above 30\,keV, when the advantage of higher transmission and larger effective aperture of Be lenses decreases. Using diamond and keeping the lens radius equal, only about half the number of lenses are required. At lower energies, diamond lenses have strong potential for applications where the lenses are subjected to high powers or intense short-duration x-ray pulse energies. Future tests of the frame-mounted diamond lens performance  when subjected to such illumination conditions will be helpful to show the predicted excellent thermal performance of diamond in white beam conditions and/or resistance to single pulse X-ray ablation.


\appendix

\section{Refractive index of diamond and beryllium}
\label{appendix:delta}

\begin{figure}
    \centering
    \includegraphics[width=0.9\columnwidth]{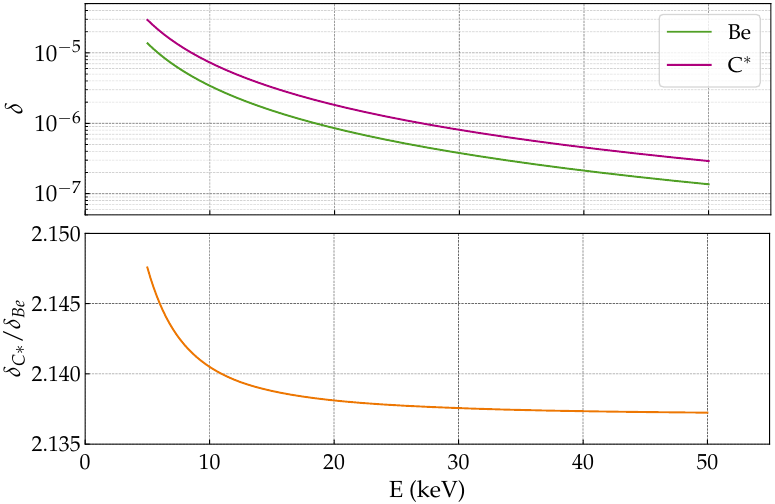}\vspace{-0.6cm}
    \caption{(top) index of refraction decrement calculated using the \textit{xraylib} library \cite{brunetti_library_2004} and (bottom) their ratio.}
    \label{fig:delta}
\end{figure}

The focal length of a stack of bi-concave x-ray lenses can be well approximated by: $f=\frac{R}{2N\delta}$, where $R$ is the radius of curvature, $N$ is the number of lenses and $\delta$ is the refractive index decrement. Figure~\ref{fig:delta} shows $\delta_{C^*}$ and $\delta_{Be}$ and their ratio for energies ranging from 5~keV to 50~keV. The mean value of this ratio is 2.14 with less than 0.5\% variation in the whole energy range. 


\ack{\textbf{Acknowledgements}}

We acknowledge the ESRF for providing beamtime on ID02, BM05 and ID06. We thank Philip Cook, Luca Capasso and Carsten Detlefs for help with setting up the beamlines BM05 and ID06. We thank Sebastian Berujon and Ruxandra Cojocaru for discussions and initial help with the XSVT data analysis code. R.C. acknowledges funding from the European Union’s Horizon 2020 research and innovation programme under grant agreement N$^{\circ}$ 101007417 within the framework of the NFFA-Europe Pilot Joint Activities. S.A. acknowledges funding from US DOE SBIR program grant N$^{\circ}$ DE-SC0013129.

\referencelist{}

\end{document}